\documentclass[eqsecnum,aps,twocolumn]{revtex4-1}
\usepackage{amsmath,amssymb,amsthm,bm,bbm}
\usepackage{psfrag,flushend}
\usepackage[dvips]{graphicx}
\usepackage[utf8]{inputenc} 
\usepackage{hyperref}

\begin{document}
\title{Comment on "Fractional topological charges and the lowest Dirac modes"} 
\author{Derar Altarawneh and Roman H\"ollwieser\footnote{Corresponding author, \email{hoellwieser@uni-wuppertal.de}}}
\affiliation {Department of Applied Physics, Tafila Technical University, Tafila , 66110 , Jordan}
\affiliation{Theoretical Particle Physics, Bergische Universit\"at Wuppertal, Gau{\ss}str. 20, 42119 Wuppertal, Germany}

\begin{abstract}
We comment on a recent article published in Phys. Rev. D98 (2018) no.9, 094513, \href{https://arxiv.org/pdf/1811.09029.pdf}{arXiv:1811.09029}, pointing out severe problems in the numerical investigation leading to questionable results and misleading conclusions during their interpretation. 
\end{abstract}


\maketitle

The author of the article under investigation (\href{https://arxiv.org/pdf/1811.09029.pdf}{arXiv:1811.09029}) claims to introduce vortex configurations with fractional topological charges and analyzes the behavior of fundmental and adjoint fermions in their background, using both, the overlap and asqtad staggered fermion formulations. We show that the considered vortex configurations contain singularities, {\it i.e.}, in lattice terms, maximally non-trivial plaquettes, which spoil the admissibility~\cite{Luscher:1981zq} of the gauge field configurations, making them non-suitable for lattice fermion investigations. In an attempt to remove the singularities using simple smearing or cooling, we find that the configurations are meta-stable and vanish or transform into a double instanton configuration, resolving an obvious discrepancy in the fermionic zero modes presented in the article under investigation but ignored by its author.

The original planar vortices parallel to two of the coordinate axes in $SU(2)$ lattice gauge theory were introduced in~\cite{Jordan:2007ff} and analyzed in detail in~\cite{Hollwieser:2011uj}. Lattice gauge links of a particular space-time direction, {\it e.g.} $x$, in a single perpendicular 3D lattice slice (in our example $yzt$) rotating gradually from $0$ to $\pi$ and on to $2\pi$ or back to $0$ in a particular subgroup of $SU(2)$, {\it e.g.} $\sigma_1$, while crossing the 3D slice in one of its directions, {\it e.g.} $y$, create two vortex sheets parallel to the remaining two space-time directions (in our example the $zt$-plane) forming a closed vortex surface due to periodic boundary conditions. The crossing of two perpendicular parallel vortices, {\it e.g.} $xy$- and $zt$ vortices, results in four intersection points of topological charge $Q=0.5$ because of non-trivial contributions from $\vec E (zt)$ and $\vec B (xy)$ vortex plaquettes to the ${\cal F}\tilde{\cal F}$ definition of topological charge. The fractional charge contributions add up to integer values $0$ or $2$ depending on their mutual orientations due to the lattice periodicity.   

\begin{figure}[b] 
\centering
\includegraphics[width=1\columnwidth]{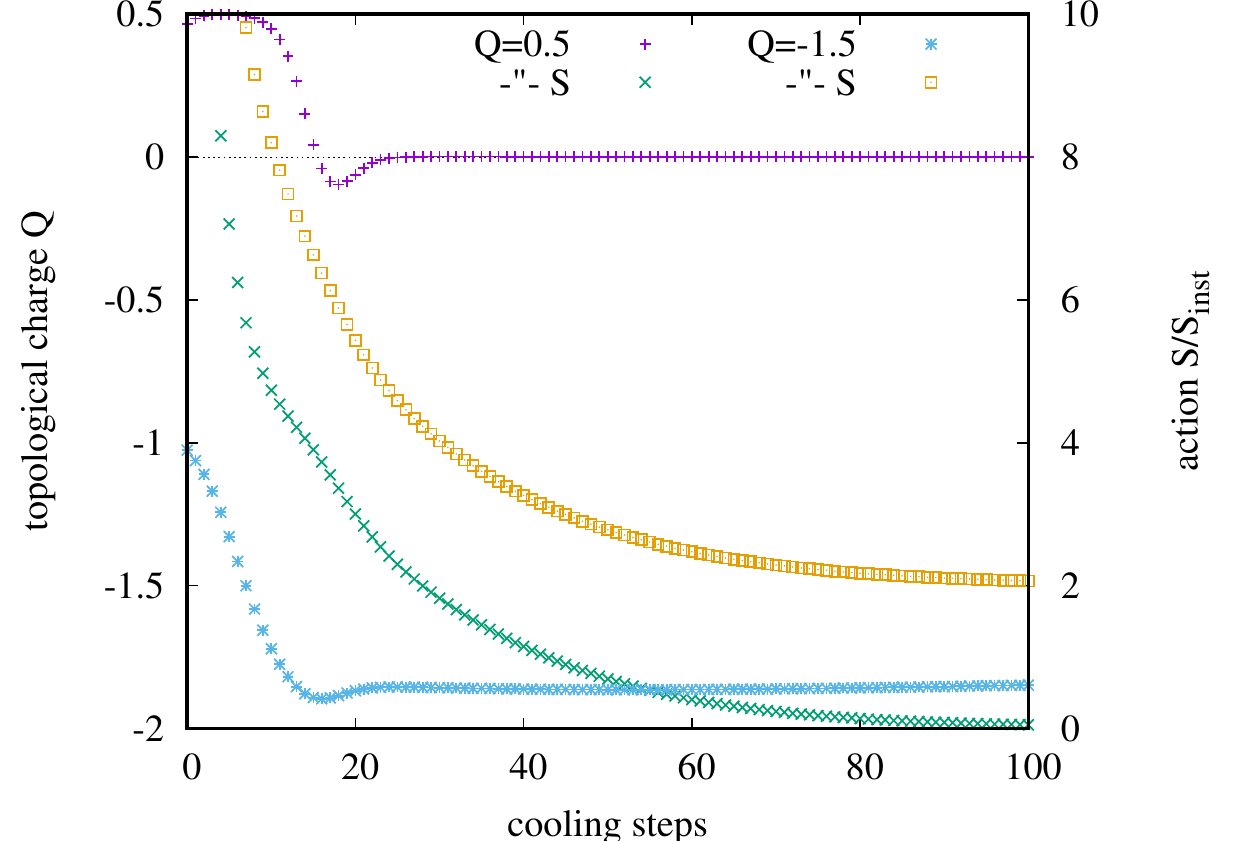}
\caption{Topological charge $Q$ and action $S$ of the $Q=0.5$ and $Q=-1.5$ configurations during cooling.}
\label{fig:cool}
\end{figure}

The idea of rotating the individual vortex sheets in different $SU(2)$ subgroups was actually explored in~\cite{Hollwieser:2012kb} already, reporting the obvious shortcomings of this approach. The motivation is to remove three of the four topological charge contributions due to the fact that crossings of $\vec E (zt)$ and $\vec B (xy)$ plaquettes with different $SU(2)$ color directions $\sigma_k$ and $\sigma_l$ ($k\neq l$) give no (real) topological charge contribution, being left with a fractional topological charge $Q=0.5$ configuration from a single vortex intersection as shown in Fig. 2c  in~\cite{Hollwieser:2012kb}. The downside however is that this introduces maximally non-trivial plaquettes $P\approx\mathrm{i}\sigma_k\mathrm{i}\sigma_l\mathrm{i}\sigma_k\mathrm{i}\sigma_l=\mathrm{i}\sigma_m\mathrm{i}\sigma_m=-\mathbbm{1}$ ($k\neq l\neq m$) from non-trivial vortex links in different $SU(2)$ subgroups in the vortex sheets. The configurations considered in \href{https://arxiv.org/pdf/1811.09029.pdf}{arXiv:1811.09029} in fact show three maximally non-trivial plaquettes and high action density peaks similar to the ones presented in Fig. 2d in~\cite{Hollwieser:2012kb}. The $Q=0.5$ configuration is in fact meta-stable and vanishes after a few cooling or smearing steps as shown in Fig.~\ref{fig:cool}. 

\begin{figure}[b] 
\centering
\includegraphics[width=0.51\columnwidth]{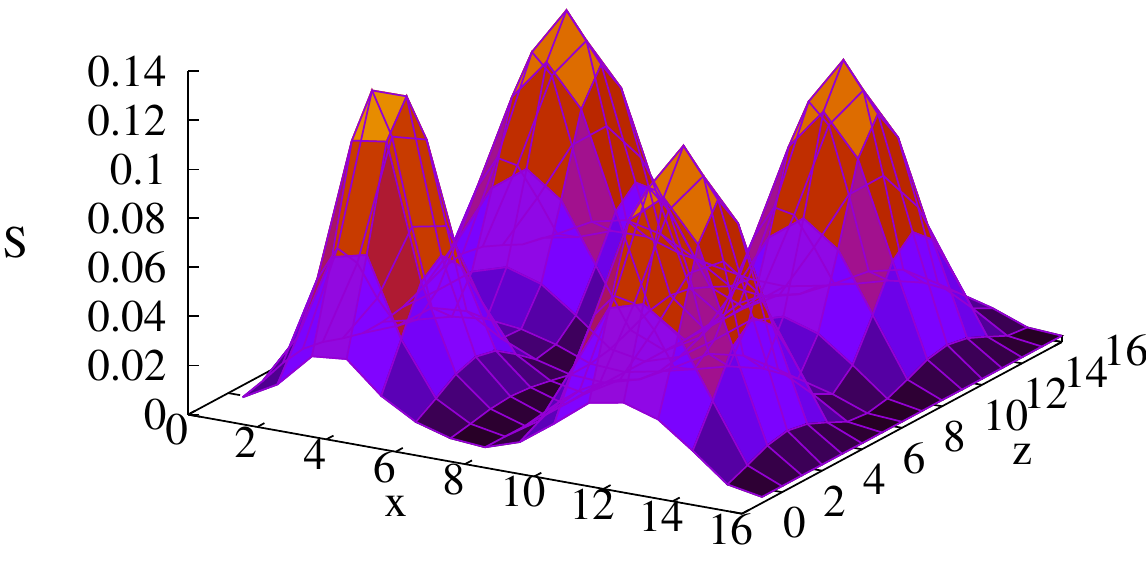}\includegraphics[width=0.51\columnwidth]{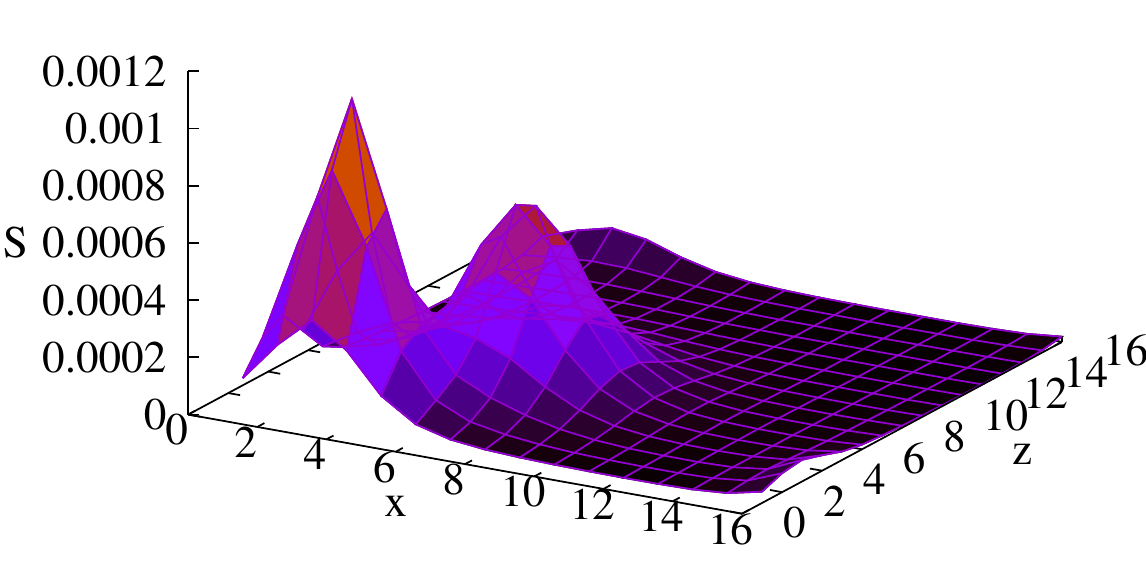}
\includegraphics[width=0.51\columnwidth]{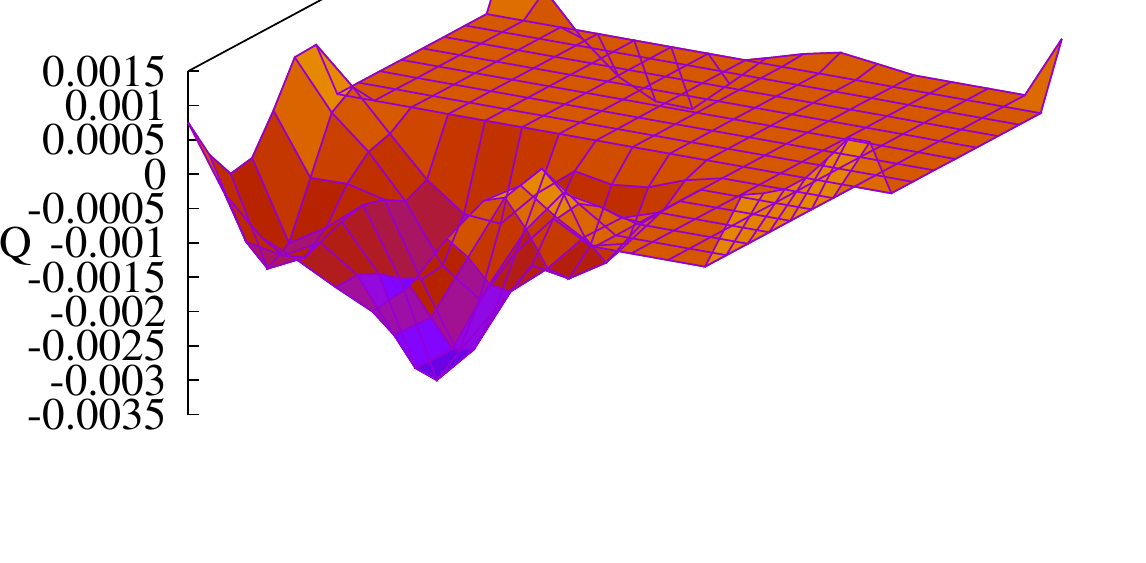}\includegraphics[width=0.51\columnwidth]{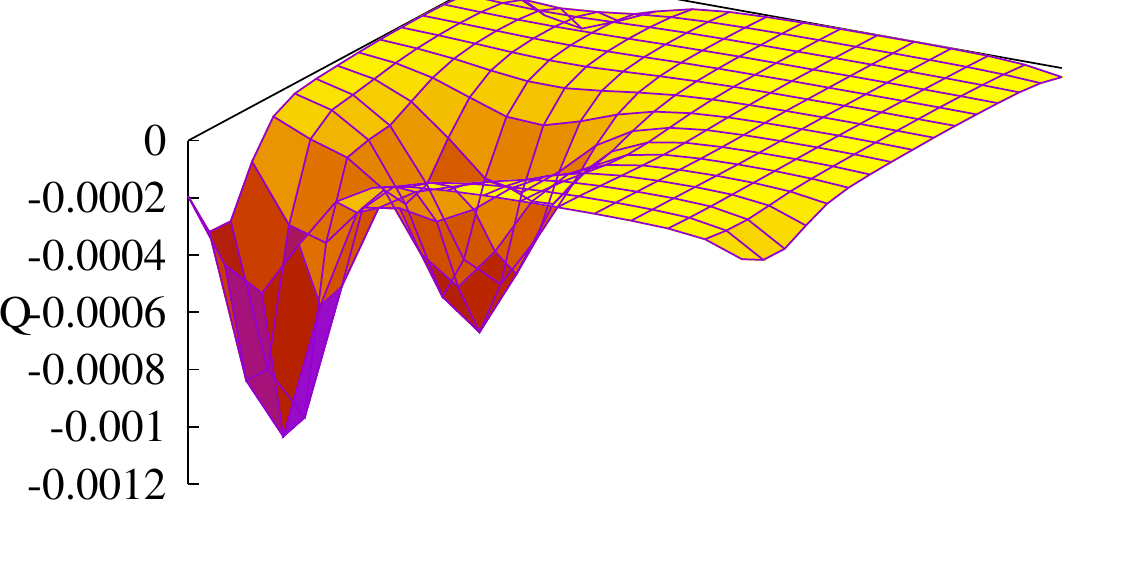}
\vspace{-6mm}
\caption{Action (top) and topological charge (bottom) density maxima of the $Q=1.5$ configuration at zero (left) and 100 (right) cooling steps in the $xz$ intersection plane resp. parallel boundary plane for the top left plot.}
\label{fig:atdens}
\end{figure}

Now, adding a so-called colorful plane vortex as introduced in~\cite{Nejad:2015aia}, {\it i.e.}, adding a non-trivial spherical color structure similar to the colorful spherical vortex introduced in~\cite{Jordan:2007ff}) around the non-trivial topological charge contribution, does not resolve the problem. What happens is that the single $Q=0.5$ contribution flips its sign and the color structure adds another $Q=-1$ contribution resulting in what is denoted the $Q=-1.5$ configuration. However, the singular nature due to the remaining three vortex intersection with artificial $Q=0$ contributions remains and even smoothing the thin colorful plane vortex only gauge transforms the non-trivial plaquettes to the lattice boundary, showing up as action maxima in the top left plot of Fig.~\ref{fig:atdens}. During cooling this configuration interestingly turns into two separate anti-instantons, totaling in topological charge $Q=-2$ and two instanton actions $S=2S_{inst}=16\pi^2$ as shown in Fig.~\ref{fig:cool} with action and topological charge densities presented in Fig.~\ref{fig:atdens} (right column). This seems to explain the two (four) fundamental zero modes of the overlap (staggered) Dirac operator which cause an obvious discrepancy with the index theorem~\cite{Atiyah:1971rm} for the supposedly $Q=-1.5$ configuration.  

However, the main criticism to be made clear is that neither the $Q=0.5$, nor the $Q=-1.5$ configurations are nearly smooth enough to calculate fermionic eigenmodes. The maximally non-trivial ($P=-1$) plaquettes certainly violate the admissibility condition~\cite{Luscher:1981zq} requiring $1-P\leq0.015$. Therefore it is absolutely improper to discuss the numbers of zero modes, let alone interpret the low-lying spectrum of the Dirac operator. Even though the adjoint zero modes seem to give reasonable results, the numbers of fundamental modes for the $Q=-1.5$ configuration are apparently wrong, a fact that is completely ignored in the discussion. Instead there is a lengthy and rather confusing discussion of the low-lying Dirac eigenmodes obtained from these singular configurations, and even some attempt of a scaling study and "the continuum limit of the background configurations". 

Because of the singular nature of the considered gauge field configurations we believe that these numerical studies are rather inconclusive, however we can at least agree on the final conclusion, that these configurations play no role for SCSB, nor in the QCD vacuum in the first place, since due to their high action they are highly suppressed in the path integral. 

This should not harm the important role of center vortices in the QCD vacuum in general though, proofing crucial for quark confinement and chiral symmetry breaking and we want to stress that the results in~\cite{hollwieser:2008tq,Hollwieser:2009wka,Hollwieser:2010mj,Hollwieser:2010zz,Hollwieser:2011uj,Schweigler:2012ae,Hollwieser:2012kb,Hollwieser:2013xja,Hoellwieser:2014isa,Hollwieser:2014mxa,Faber:2014cya,Hollwieser:2014lxa,Greensite:2014gra,Hollwieser:2015gra,Nejad:2015aia,Hollwieser:2015koa,Hollwieser:2015qea,Altarawneh:2015bya,Altarawneh:2016ped,Faber:2016bjg,Hollwieser:2017xmn,Faber:2017alm,Golubich:2018ubu} are not affected by the above criticism.

\vspace{-6mm}

\bibliographystyle{unsrt}
\bibliography{paper}

\end{document}